\documentclass[12pt,a4paper]{article}

\usepackage{amsmath}
\usepackage{amsfonts}
\usepackage{amssymb}
\usepackage{amsthm}

\input{epsf}             

\newcommand{\R}{\mathbb R}              

\newcommand{\dd}{{\rm d}} 

\theoremstyle{plain} 

\theoremstyle{definition}

\theoremstyle{remark}

\begin{document}

\title{Feynman diagams coupled to three-dimensional quantum gravity}

\author{John W. Barrett
\thanks{Copyright \copyright\ John W. Barrett 2005}
\\ \\
School of Mathematical Sciences\\
University of Nottingham\\
University Park\\
Nottingham NG7 2RD, UK\\
\\
john.barrett@nottingham.ac.uk}

\date{}

\maketitle

\begin{abstract}  A framework for quantum field theory coupled to three-dimensional quantum gravity is proposed. The coupling with quantum gravity regulates the Feynman diagrams. One recovers the usual Feynman amplitudes in the limit as the cosmological constant tends to zero. 
\end{abstract}



The purpose of this note is to give further detail of the proposal to define a quantum field theory of particles coupled to three-dimensional quantum gravity which was developed in \cite{B,B2}, using technical developments and ideas from \cite{FL1,GI,FL2,BGM}. 

There are two ways of formulating the usual Feynman diagram in flat space - the position representation and the (more common) momentum representation. Both of these can be used, after a little reformulation, to give the same theory of the coupling of Feynman diagrams to quantum gravity. I will describe each of these in turn.

\subsection*{The position representation}
In the position space representation of a Feynman amplitude in quantum field theory, one integrates over a position variable $x\in\R^3$ for each vertex of the diagram. The integrand is a product of Feynman Green functions $G_F(|x_i-x_j|)$ for each edge, together with (for a scalar field theory) a coupling constant $g$ at each vertex.
\begin{equation} 
I=g^{\text{\# vertices}}\int \prod_{ij}  G_F(|x_i-x_j|)\prod_k \dd x_k \label{posnfd}\end{equation}
The notation for the product $\prod_{ij}$ means sum over all edges in the graph, $i$ and $j$ labelling the vertices at the end of each edge.
The Green function $G_F$ actually depends only on the distance between the vertices and so one can reformulate this integral using the distances 
$$ r_{ij}=|x_i-x_j|$$
along the edges of the graph as the integration variables,
\begin{equation}
I=g^{\text{\# vertices}}\int \mu(r_{12},\ldots)\prod_{ij}  G_F(r_{ij}) \dd r_{ij}\label{distfd}\end{equation}
where the function $\mu$ of all of the distances $r_{12},\ldots$ is the Jacobian factor for the change of variables from the $x_i$ to the $r_{ij}$. (Note, the notation $r_{12},\ldots$ will be taken to mean the list of all such variables for all edges in the graph.)
 In particular $\mu$ is zero whenever the triangle inequalities (and their generalisations to polygons) for edges around a closed circuit are violated.
The proposal for coupling the Feynman diagram $\Gamma$ to three-dimensional quantum gravity is to replace the variables $r_{ij}$ by discrete spin variables $\rho_{ij}\in\{0,1/2,1,\ldots,(r-2)/2\}$ for an integer $r$ (which determines the cosmological constant by $\Lambda=1/r^2$).
The kinematical factors $\mu(r_{12},\ldots)$ are replaced by the quantum gravity partition function $Z(M,\Gamma(\rho_{12},\ldots))$ of \cite{B}, $M$ being the three-dimensional space-time manifold. This is a relative version of the Turaev-Viro partition function of $M$, where the spin variables of the Turaev-Viro partition function are fixed to have the given values $\rho_{ij}$ along the edges of the graph $\Gamma$. The integration 
$$\int\prod_{ij}\dd r_{ij}$$
is then replaced by the summation
$$\sum_{\rho_{12},\ldots}$$ In this replacement there is actually a shift by the constant $1/2$, so that distance $r$ corresponds to the discrete variable $\rho+1/2$.

The form of the appropriate Feynman propagator (and the coupling constant)
is not fixed by quantum gravity; in a Lagrangian formulation these would be determined by the matter part of the Lagrangian. So the best one can say is that $G_F(r)$ has to be replaced by a function $G_D(\rho_{ij})$ on a discrete set, which reduces to it in the $r\to\infty$ limit,
$$G_D(\rho+1/2)\to G_F(\rho).$$

The formula for the proposal is then
\begin{equation}\label{proposal}I_{QG}=g^{\text{\# vertices}}\sum_{\rho_{12}\ldots}Z(M,\Gamma(\rho_{12},\ldots))
\prod_{ij}G_D(\rho_{ij}).\end{equation}
which is finite because it is a finite sum.

For sufficiently simple Feynman graphs, this proposal reduces to the previous integral expression (\ref{posnfd}) in the limit $r\to\infty$. For example, it is shown in \cite{B} that if the graph $\Gamma$ is a triangle (topologically the unknot) then 
 \begin{equation} \label{triangle} Z(M, \Gamma(\rho_{12},\rho_{23},\rho_{31}))=
C  \sin \frac \pi r (2\rho_{12}+1) \; \sin \frac \pi r (2\rho_{23} +1)\; \sin \frac \pi r (2\rho_{31}+1).
 \end{equation}
if $\rho_{12}+\rho_{23}+\rho_{31}$ is an integer (rather than half-integer) and there exists a geodesic triangle on the sphere of radius $r/2\pi$ with edge lengths $\rho_{12}+1/2,\rho_{23}+1/2,\rho_{31}+1/2$; otherwise $Z=0$.

This formula is obviously the discrete analog of the continuum formula
\begin{equation}\mu_{S^3}(r_{12},r_{23},r_{31})=
 C \sin \frac {2\pi r_{12}} r \;\sin \frac {2\pi r_{23}}r \;\sin \frac {2\pi r_{31}}r 
 \end{equation}
if $(r_{12},r_{23},r_{31})$ are the sides of a spherical triangle, and $0$ otherwise. 
This second formula is a good approximation to the first when the spacing between the discrete variables can be considered small, i.e., when the value of $r$ is large. It is the formula for the distribution function for the edges lengths when the vertices of a triangle are integrated over the three-sphere, and so $\mu_{S^3}$ is the spherical analog of $\mu$. Therefore $Z(M, \Gamma(\rho_{12},\rho_{23},\rho_{31}))$ converges  to the corresponding flat space factor $\mu$ of (\ref{distfd}) in the limit $r\to\infty$, with appropriate scalings. This result holds for any polygonal graph, and one would conjecture that it is also true for any planar graph. Note that the relevant triangle inequalities hold automatically due to the properties of the quantum gravity partition function. The reason the result works is that three-dimensional quantum gravity is dominated by the classical solutions to the Einstein equations, which are metrics which are all locally diffeomorphic to $S^3$ with radius $r/2\pi$ (and so for the $r\to\infty$ limit, flat space). So the theory behaves like a quantum version of flat space.

For non-planar graphs the quantum gravity formulation differs from the flat-space formulation of Feynman diagrams because one part of the graph responds to the gravitational effect of the other. For example, for the Hopf link, the particle moving around one component of the link sees the conical singularities induced in the (fluctuating) metrics through the middle of it by the mass of the other particle. Also, if a particle follows a knotted trajectory then the result differs from an unknotted trajectory due to the self-interaction of the particle via its own gravitational field. This phenomenon was explored in \cite{B2}. These results are due to the use of one-dimensional particles in a three-dimensional space-time for which it is well-known that a Chern-Simons type theory will give knotting and linking phenomena. The particles are a type of non-abelian anyon.

\subsection*{The momentum representation}
The proposal can be formulated equivalently in the momentum picture. The Feynman diagram amplitude (\ref{posnfd}) can be written in terms of the momentum variable $p\in\R^3$ on each edge of the graph, using a Fourier transform of the Feynman propagator
\begin{equation} 
I=g^{\text{\# vertices}}
\int\prod_i\delta(\sum_jp_{ij})\prod_{lm} \widetilde G_F(|p_{lm}|)\dd p_{lm} \label{momfd}\end{equation}
Again, the propagator only depends on the length of the virtual momentum,
$$k_{ij}=|p_{ij}|$$
which will be called the virtual mass of the particle. This must be distinguished carefully from the on-shell mass $m$ of a particle, which just appears as a parameter in the propagator
$$\widetilde G_F(k)=\frac1{k^2+m^2}.$$
The integral can then be reformulated as
$$I=g^{\text{\# vertices}}\int \mu'(k_{12},\ldots)\prod_{ij} \widetilde G_F(k_{ij}) \dd k_{ij}.$$
The proposal for coupling to gravity in this picture is to replace the $k_{ij}$ by discrete variables $\kappa_{ij}\in\{0,1/2,1,\ldots,(r-2)/2\}$ and the $\mu'$ with a second quantum gravity partition function $Z_R(M,\Gamma(\kappa_{12},\ldots))$, also introduced in \cite{B}. The $Z_R$ is defined by a discrete Fourier transform using the Fourier kernel
$$F_\kappa(\rho)=(-1)^{2\kappa} \frac{\sin \frac{\pi}{r} (2\rho+1)(2\kappa+1)}{\sin \frac{\pi}{r}(2\rho+1)}$$
 by
$$Z_R(M,\Gamma(\kappa_{ij}))= \sum_{\rho_{12},\ldots}Z(M,\Gamma(\rho_{12},\ldots))\prod_{ij}F_{\kappa_{ij}}(\rho_{ij})
 $$ 
 Thus in the momentum representation the proposal is to couple the Feynman diagram to quantum gravity by the formula
 $$I_{QG}=g^{\text{\# vertices}}\sum_{\kappa_{12}\ldots}Z_R(M,\Gamma(\kappa_{12},\ldots))
\prod_{ij}\widetilde G_D(\kappa_{ij}).$$
This agrees with the position representation formula (\ref{proposal}) if the corresponding propagator fomulae are related by the Fourier transform
\begin{equation}\label{propagator}G_D(\rho)=\sum_\kappa\widetilde G_D(\kappa)F_\kappa(\rho). \end{equation}
In this formulation $Z_R$ gives the correct coupling of particles with specified virtual masses to quantum gravity. As discussed in \cite{B}, the correct triangle inequalities for the $\kappa_{ij}$ meeting at the $i$-th vertex hold, and the quantum gravity partition function automatically incorporates the conservation of energy-momentum of the propagating particles.

\subsection*{External lines}
In quantum field theory in flat space, Feynman diagrams with external lines determine a function of the coordinates of the ends of the external lines. When coupled to quantum gravity the coordinates lose their invariant meaning, so one has to reformulate these amplitudes in terms of closed diagrams. For example, the graph
$$\epsfbox{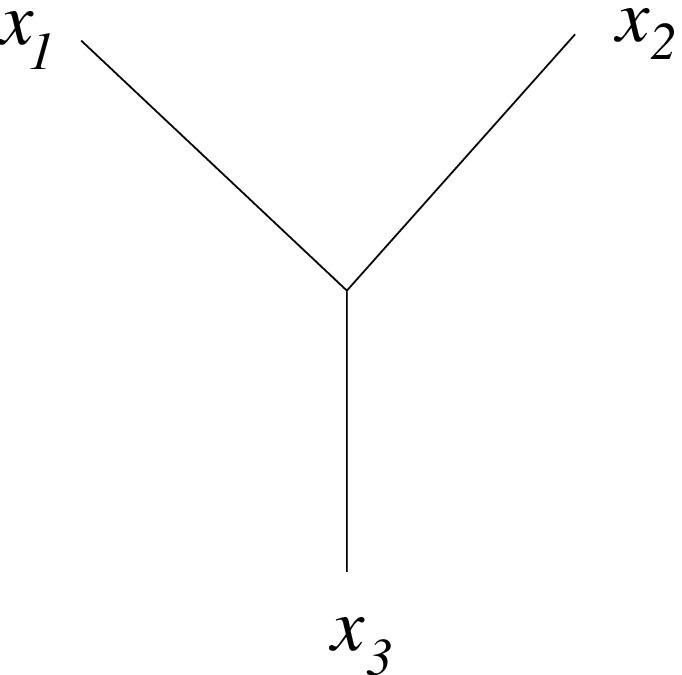}$$
is a function of the three position vectors $x_1$, $x_2$, $x_3$. However this function depends only on the invariant distances between $x_1$, $x_2$ and $x_3$. Accordingly, this amplitude can be reconstructed from the closed graph amplitude determined by the tetrahedral graph $\Gamma$
$$\epsfbox{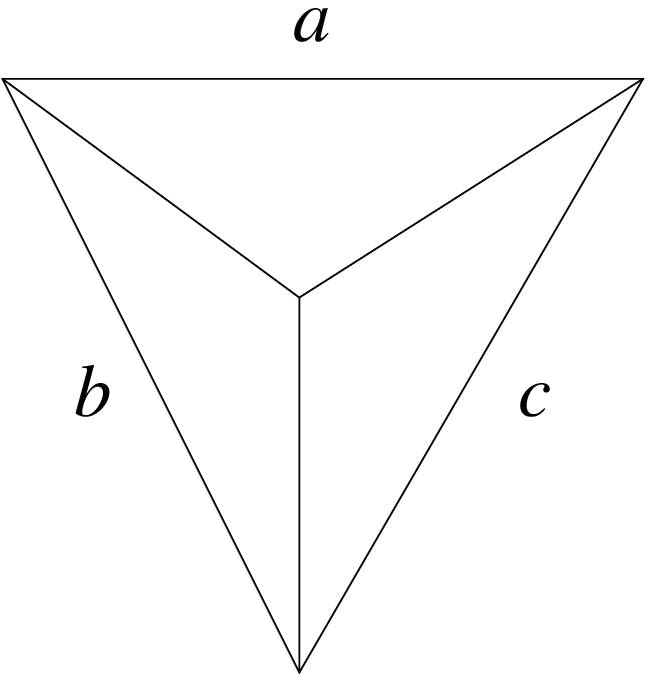}$$
in which the distances along the outer edges are fixed at values $a$, $b$ and $c$. The calculation of this function is therefore
$$\psi(a,b,c)=\sum_{\rho_{14}\rho_{24}\rho_{34}}
Z(M,\Gamma(a,b,c,\rho_{14},\rho_{24},\rho_{34})G_D(\rho_{14})G_D(\rho_{24})G_D(\rho_{34})$$
the additional $\rho$ variables labelling the other three unmarked edges in the diagram. This amplitude can be evaluated in terms of sums of products of two $q-6j$ symbols.

The simplest case of this procedure is the particle propagator between two points. Suppose these two points are a distance $a$ apart. Then the corresponding closed diagram
$$\epsfbox{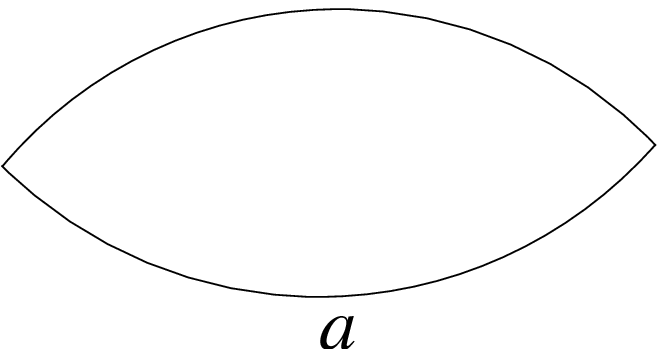}$$
has two lines between the two points, and the distance along one of them is fixed as $a$. The amplitude is then
$$\psi(a)=\sum_{\rho}
Z(M,\Gamma(a,r))G_D(\rho).$$
The partition function is 
$$Z(M,\Gamma(a,\rho))=C' \delta_{a\rho}\sin^2\frac\pi r(2a+1)$$
Therefore 
$$\psi(a)=C' G_D(a) \sin^2\frac\pi r(2a+1),$$
from which one obtains the correct Green function $G_D(a)$ on dividing by the normalising factor $\sin^2\frac\pi r(2a+1)$, which is just a constant times the area of the sphere of radius $a$ on $S^3$. One can also analyse the contribution of the virtual mass $\kappa$ to this amplitude by taking the observable where $\kappa$ is also fixed on the unmarked line in the above diagram. The corresponding amplitude with the same normalisation is, from (\ref{propagator}), proportional to
$$ \phi(a)=\widetilde G_D(\kappa)F_\kappa(a)$$
which, as remarked in \cite{B}, obeys the Helmholtz equation
$$\nabla^2\phi= \frac{-16\pi^2}{r^2}\kappa(\kappa+1)\phi$$
 on $S^3$. In the limit $r\to\infty$ it will obey the corresponding equation in flat space.

Some aspects of the theory have not been addressed in the above constructions. Most particularly, the systematic generation of the Feynman amplitudes coupled to gravity, and with it the symmetry factors.  


\begin{thebibliography}{99}
\bibitem[B] {B}  Barrett, J.W. Geometrical measurements in three-dimensional quantum gravity.  gr-qc/0203018  Internat. J. Modern Phys. A18  (2003),  97--113.
\bibitem[BGM]{BGM} Barrett, J.W.; Garcia-Islas, J.M.; Faria Martins, J. Observables in the Turaev-Viro and Crane-Yetter models. math.QA/0411281
\bibitem[B2] {B2} Barrett, J.W. Feynman loops and three-dimensional quantum gravity. gr-qc/0412107. To appear in Int. J. Mod Phys. A.     
\bibitem[TV]{TV} Turaev, V.G.; Viro, O.Ya. State sum invariants of $3$-manifolds and quantum $6j$-symbols.  Topology  31  (1992)  865--902
\bibitem[FL1]{FL1} Freidel, L.; Louapre, D.   
    Ponzano-Regge model revisited I: Gauge fixing, observables and interacting spinning particles. hep-th/0401076.
\bibitem[FL2]{FL2} Freidel, L.; Louapre, D. Ponzano-Regge model revisited II: Equivalence with Chern-Simons. gr-qc/0410141.
\bibitem [GI]{GI} Garcia-Islas, J.M.   Observables in 3-dimensional quantum gravity and topological invariants. gr-qc/0401093   Class.Quant.Grav. 21 (2004) 3933-3952 
\end{thebibliography}
\end{document}